\documentclass[galaxies,article,pdftex,moreauthors]{Definitions/mdpi}

\makeatletter
\let\linenumbers\relax

\let\nolinenumbers\relax
\makeatother
\firstpage{1} 
\pubvolume{1}
\issuenum{1}
\articlenumber{0}
\pubyear{2026}
\copyrightyear{2026}
\datereceived{ } 
\daterevised{ } 
\dateaccepted{ } 
\datepublished{ } 

\usepackage[flushleft]{threeparttable}
\usepackage{comment}
\usepackage{xcolor}

\defcitealias{2021MNRAS.502.5686I}{IB21}
\defcitealias{2023ApJ...944...88L}{L23}
\defcitealias{2023AA...674A..18C}{C23}
\defcitealias{2025MNRAS.536.2749M}{Mv25}
\defcitealias{1989ApJ...345..245C}{CCM89}
\newcommand{\msun}{M_{\odot}}
\newcommand{\apj}{ApJ}

\newcommand{\apjs}{ApJS}
\newcommand{\aap}{A\&A}
\newcommand{\mnras}{MNRAS}
\newcommand{\aj}{AJ}
\newcommand{\actaa}{Acta Astron.}
\newcommand{\pasp}{PASP}
\newcommand{\apss}{Ap\&SS}

\Title{Photometric metallicity of Galactic RR Lyrae stars in the Gaia~DR3 era}

\Author{Mahiguhappriya Prakash $^{1,2}$\orcidA{}, Susmita Das $^{3,*}$\orcidB{}, Harinder P. Singh $^{1}$\orcidC{} and Nitesh Kumar $^{4}$\orcidD{}}

\AuthorNames{Mahiguhappriya Prakash, Susmita Das, Harinder P. Singh and Nitesh Kumar}

\address{%
$^{1}$ \quad Department of Physics and Astrophysics, University of Delhi, North Campus, Delhi-110007, Delhi, India\\
$^{2}$ \quad Indian Institute of Astrophysics (IIA), II Block, Koramangala, Bengaluru-560034, Karnataka, India\\
$^{3}$ \quad Inter-University Center for Astronomy and Astrophysics (IUCAA), Post Bag 4, Ganeshkhind, Pune-411007, Maharashtra, India\\
$^{4}$ \quad Department of Physics, Applied Science Cluster, University of Petroleum and Energy Studies (UPES), Dehradun-248007, Uttarakhand, India}

\corres{Correspondence: susmita.das@iucaa.in}

\abstract{RR~Lyrae stars are pulsating variables crucial for distance determination and galactic structure studies. Metallicities of fundamental-mode (RRab) RR Lyrae stars are commonly derived from photometry using empirical relations involving the Fourier parameter $\phi_{31}$ and the pulsation period. We present a new, calibrated $G$-band relationship between pulsation period $P$, Fourier parameter $\phi_{31}$, and metallicity [Fe/H] for galactic RR~Lyrae stars from the \textit{Gaia} survey. A set of 72 fundamental mode RR~Lyrae stars were identified for deriving the relation in the $G$-band, after visual examination of their light curves. Unlike recent large-scale calibrations, our relation prioritizes calibration purity by anchoring exclusively to a homogeneously analyzed sample of high-resolution spectroscopic metallicities from the literature. Our best fit relation is $\text{[Fe/H]} = (-6.93 \pm 0.58) - (6.04 \pm 0.37)P + (1.65 \pm 0.11)\phi_{31}$. We compare the [Fe/H] predicted by our relation for the stars in our calibration sample with that obtained from previously established relations in the $G$-band using different approaches. Our calibrated $G$-band $P$-$\phi_{31}$-[Fe/H] relationship demonstrates high reliability when validated against spectroscopic data, achieving a negligible bias of $0.00$ dex and an empirical RMS scatter of 0.26 dex. Furthermore, by applying an Orthogonal Distance Regression (ODR) routine that fully propagates parameter covariance, we establish a mathematically strict empirical baseline whose theoretical uncertainties perfectly align with this observed dispersion. We find that the inclusion of the $R_{21}$ Fourier parameter offers no significant improvement in metallicity estimation. Comparisons with literature confirm that our linear relation aligns closely with other \textit{Gaia} DR3-based studies, while offering improved precision over older DR2-based relations.}

\begin{document}
\nolinenumbers

\section{Introduction}\label{introduction}

RR~Lyrae stars (RRLs) are low-mass (0.5--0.8\,$\msun$) Population II variables situated at the intersection of the horizontal branch and the classical instability strip of the Hertzsprung-Russell Diagram \citep{2004ApJS..154..633C}. Exhibiting pulsation periods between 0.2 and 1.2 days with photometric amplitudes of $\leq$2 mag, their variability is driven by the $\kappa$--mechanism, which arises from opacity changes within the He II partial ionization zone \citep{2001MNRAS.326.1183B}. Consequently, they serve as robust tracers of old stellar populations in the Milky Way (halo, bulge, and disc) as well as in galaxies across the Local Group \citep{1985MNRAS.216..873L, 2006MNRAS.372.1675S, 2015ApJ...799..165B, 2015ApJ...807..127M, Beaton_2018, 2020AJ....160..220B}.

Accurate determination of metal abundances in stellar sources traditionally demands spectroscopic analysis. However, acquiring such data requires significant observational resources and time, with the resolving power of the instrument acting as a major limiting factor \citep{Zhang_2020}. While massive spectroscopic surveys utilizing instruments like the Large Sky Area Multi-Object Fiber Spectroscopic Telescope (LAMOST; \citealt{2012RAA....12.1197C}) can collect thousands of spectra in a single exposure, they still struggle to probe faint sources at very large distances or in regions of high extinction. In contrast, photometric surveys offer a significantly higher observational efficiency, yielding vast amounts of data across deeper volumes in less time \citep{2025AA...693A.306F}. For fundamental-mode RRLs (RRab), photometric light curves encode crucial information regarding their intrinsic metal abundances \citep{2018MNRAS.481.2000D}. Therefore, establishing a mathematical relationship between metallicity ([Fe/H]), pulsation period, and light curve morphology offers a powerful, efficient alternative for estimating the chemical properties of these pulsating variables.

Early work by \citet{1996AA...312..111J} demonstrated that the shape of the light curves of RRab stars at optical wavelengths can be related to their metal abundance. Their work presents a linear relationship connecting the metallicity [Fe/H] of the star with its pulsation period $P$ and the lower-order Fourier parameters ($\phi_{31}$) obtained from the Fourier decomposition of its light curve in the $V$-band. Subsequent studies by \citet{article, 2013ApJ...773..181N, 2016CoKon.105...53M, 2016ApJS..227...30N, 2021MNRAS.502.5686I, 2021ApJ...912..144M} extended this analysis to well-sampled RRab light curves obtained from various survey programs, such as the Kepler Space Telescope (in $K_p$-band; \citet{2010Sci...327..977B}), the $I$-band Optical Gravitational Lensing Experiment \citep{2015AcA....65....1U}, and the $R$-band Palomar Transient Factory \citep{2009PASP..121.1395L}. Building upon this legacy, recent studies by \citet{2023AA...674A..18C} and \citet{2025MNRAS.536.2749M} have pioneered the application of these photometric metallicity calibrations to the latest generation of broad-band, space-based photometry from the \textit{Gaia} mission.

Determining accurate metallicities for these stars is crucial because they obey well-defined period-luminosity ($PL$) relations, especially at longer wavelengths \citep{Bhardwaj_2021, Zgirski_2023, Bhardwaj_2023}. As metallicity effects introduce scatter into these $PL$ relations, precise [Fe/H] estimates are required to refine RRLs as distance indicators \citep{refId0}. The most recent large-scale surveys provide massive, high-precision photometry in mission-specific passbands, most notably the \textit{Gaia} $G$-band. The European Space Agency's \textit{Gaia} space observatory was designed to provide unprecedented astrometric and photometric precision for over a billion sources \citep{2016AA...595A...1G}. Its recent Data Release 3 (\textit{Gaia} DR3; \citet{2023AA...674A...1G}) offers extensive epoch photometry collected in three primary passbands: the broad white-light $G$-band (330--1050 nm), the blue $G_{\rm BP}$ band (330--680 nm), and the red $G_{\rm RP}$ band (630--1050 nm) \citep{Riello_2021}. The goal of the present investigation is to perform a Fourier analysis of RRL light curves to derive a new $P$-$\phi_{31}$-[Fe/H] relation specifically for the $G$-band using \textit{Gaia} DR3 photometric data. Although $G$-band and $V$-band magnitudes are closely correlated \citep{2010AA...523A..48J}, a native calibration is necessary to fully exploit the precision of \textit{Gaia} photometry.

This paper is structured as follows. In Section \ref{sec:data}, we give a brief description of the database used and the procedure for obtaining the phased light curves from the raw data. We present the Fourier decomposition followed by the $P$-$\phi_{31}$-[Fe/H] calibration in Section \ref{subsec:fourier}. In Section \ref{sec:analysis}, we present our newly derived relation for finding [Fe/H] in the $G$-band and compare the predictions of our relation with the corresponding spectroscopic [Fe/H] values. We further compare the reliability of our relation with that of previous works within the range of our calibration data in Section \ref{sec:discussion} and in Section \ref{sec:conclusion}, we present the conclusions of our study.

\begin{figure*}
    \centering
    \includegraphics[width=0.95\linewidth]{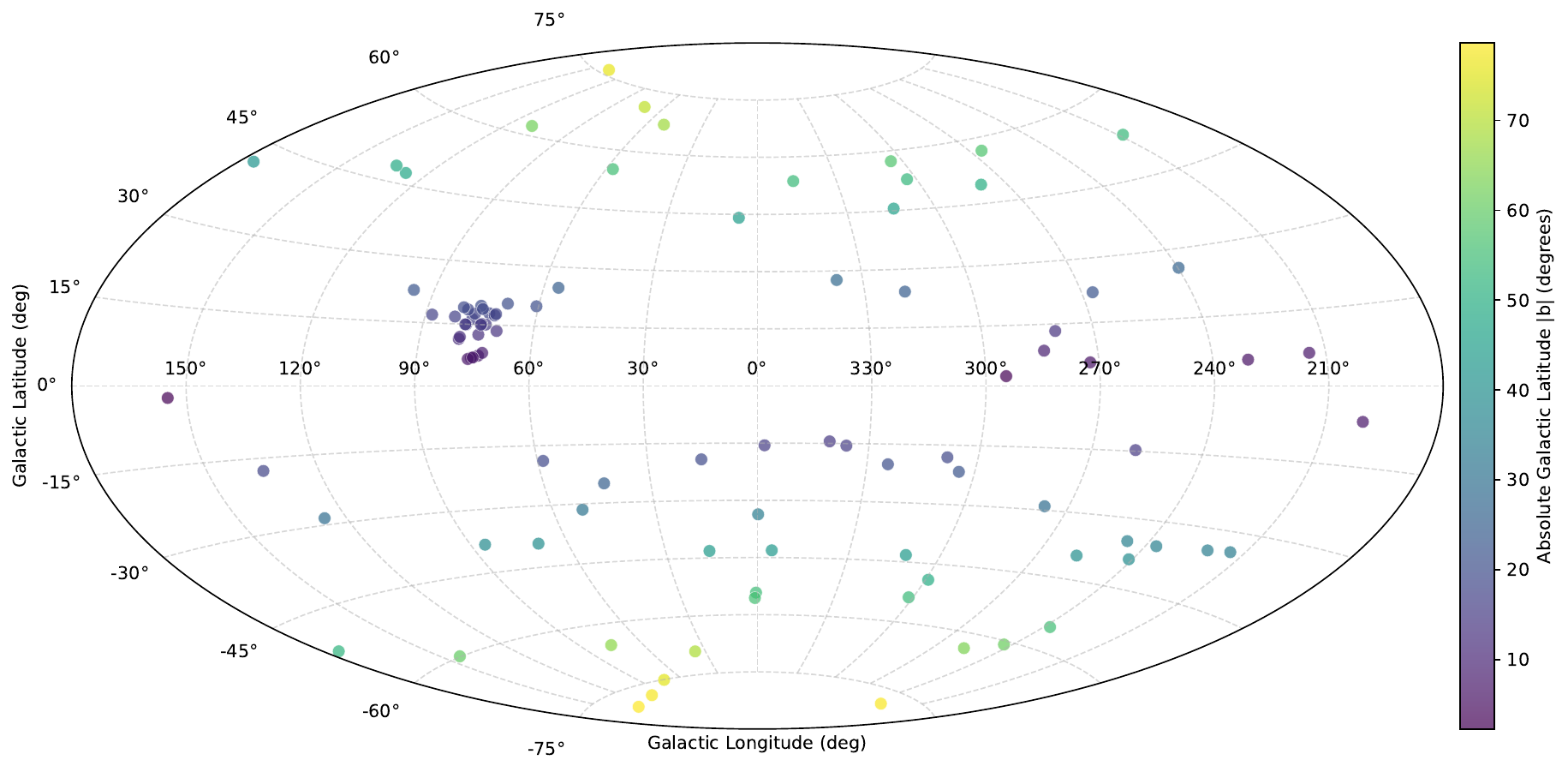}
    \caption{Galactic map of the 108 RR Lyrae (RRab) obtained from \citet{2021MNRAS.503.4719G}.}
    \label{fig:galmap}
\end{figure*}

\section{Input Data and Methods Applied}

\label{sec:data}

\subsection{The dataset}\label{subsec:dataset}
The RRLs analyzed in the present work were selected from \citet{2021MNRAS.503.4719G}, comprising 108 galactic RRab stars. To visualize the spatial distribution of this sample (Fig. \ref{fig:galmap}), we retrieved the Galactic coordinates ($l, b$) for all 108 stars via the SIMBAD astronomical database using their respective \textit{Gaia} Source IDs. The absolute Galactic latitude ($|b|$) was then calculated directly from these queried latitude values. We obtained the raw photometric data of these stars in the \textit{Gaia} passbands $G, G_{\rm BP}, G_{\rm RP}$ from \textit{Gaia} Data Release 3 (\textit{Gaia} DR3; \citet{2016AA...595A...1G, 2023AA...674A...1G}) and utilized exclusively the $G$-band photometric time-series data for our light curve analysis. 

To account for the galactic dust extinction, we used the reddening ($E(B-V)$) values from \citet{2021MNRAS.503.4719G} for each star in our calibration sample. These were determined from the 3D maps from \citet{2019ApJ...887...93G} and further compared with the 2D reddenings from \citet{1998ApJ...500..525S}. To convert from $E(B-V)$ to the total extinction in the $G$-band ($A_G$), we assumed $R_V = 3.1$ and used the extinction reference model given by \citet{1989ApJ...345..245C} (CCM89).

Once the magnitudes are extinction corrected, the next step is to obtain the phased light curves. While initial period estimates for these sources are available in the literature \citep{2021MNRAS.503.4719G}, we re-derived the pulsation periods $P$ directly from our data using the Lomb-Scargle method \citep{1976ApSS..39..447L, 1982ApJ...263..835S}. This ensures maximum internal consistency and prevents phase smearing during the subsequent Fourier decomposition. To calculate the phase, we need an epoch reference time $t_0$. For unevenly spaced photometric data, the time of first observation is not an ideal choice. Therefore, we chose $t_0$ to be the epoch at maximum brightness (or minimum magnitude). We used Eq.~\ref{eq:phase} to determine the phase at each epoch, $t$, to obtain the final phased light curve:

\begin{equation}\label{eq:phase}
    \Phi = \text{fractional part of } \left( \frac{t - t_0}{P} \right)
\end{equation}

During the phase-folding process, anomalous data points were removed using an iterative Interquartile Range (IQR) clipping method to ensure the fidelity of the light curves. After cleaning, the final light curves contain an average of 47 photometric data points per star.

\begin{figure*}
\centering
\includegraphics[width=0.95\linewidth]{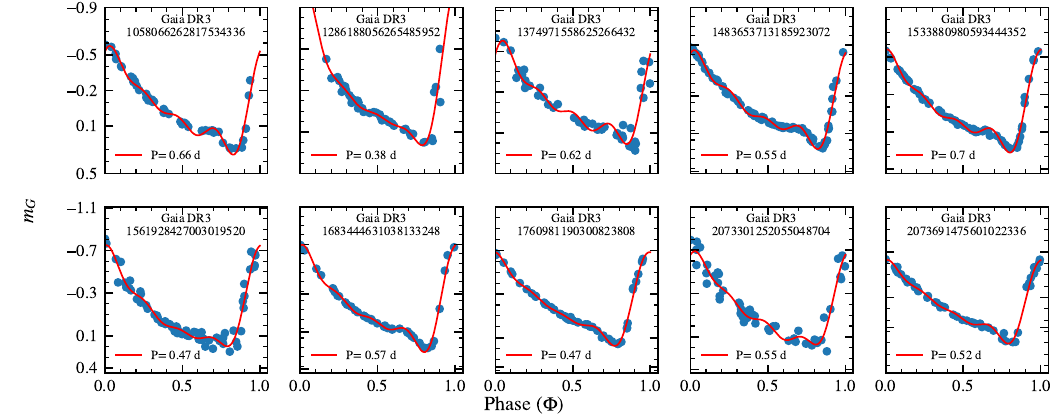}
\caption[]{Example light curves of RRab stars from $Gaia$ DR3. The outliers were removed using IQR clipping. The Fourier-fitted light curves (red solid line) are plotted over the phased light curves (blue scatter plot).}
\label{fig:lightcurves}
\end{figure*}

\begin{figure*}
    \centering
    \includegraphics[width=0.95\linewidth]{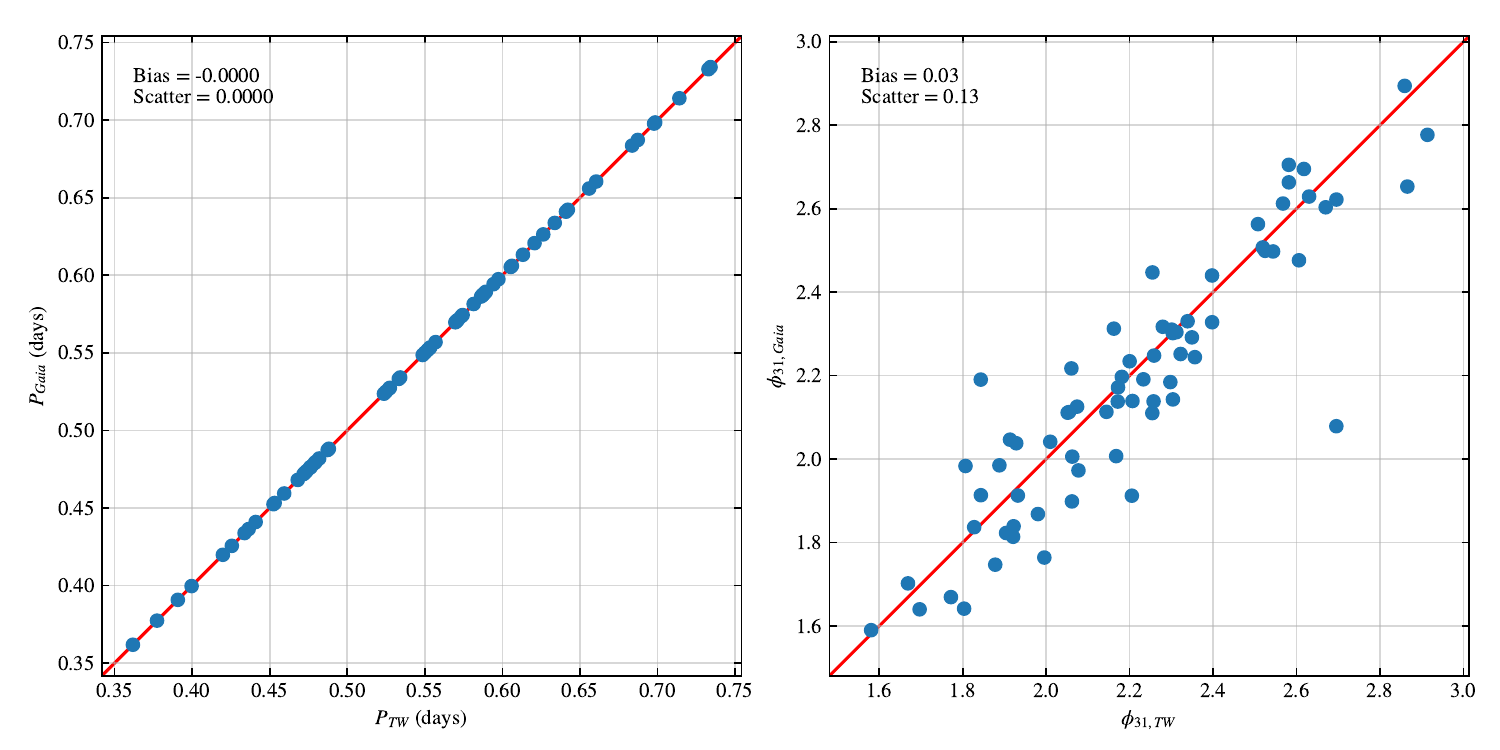}
    \caption{Comparison of the parameters derived independently in this work (TW) with the automated estimates from \textit{Gaia} DR3. \textbf{Left panel:} Pulsation periods ($P$). \textbf{Right panel:} Fourier phase differences ($\phi_{31}$). The solid red line in both panels represents the 1:1 relation. The calculated bias and RMS scatter are annotated in the upper left corner of each respective panel.}
    \label{fig:comparison_gaia}
\end{figure*}

We visually examined the light curves and selected 86 out of a total of 108 stars, rejecting those with significant gaps or insufficient data points.
{Figure}~\ref{fig:lightcurves} shows the example light curves of 10 stars from our sample. To verify the robustness of our independent frequency analysis, we compared our derived periods ($P_{\rm TW}$) against the automated period estimates published in \textit{Gaia} DR3 ($P_{\rm Gaia}$). As shown in the left panel of Figure \ref{fig:comparison_gaia}, the values are in near-perfect agreement, exhibiting no measurable bias or scatter between the two methods. This exact consensus is expected for classical pulsators, demonstrating that the standard Lomb-Scargle algorithm yields highly stable primary frequencies regardless of minor differences in pipeline implementation.

\subsection{Fourier Decomposition of light curves}\label{subsec:fourier}

Fourier decomposition is a widely used tool to quantify the shape of a periodic light curve. Previous studies have shown that the lower order terms in the Fourier expansion are sufficient to fully characterize the shape of a light curve \citep{{1981ApJ...248..291S},{1996AA...312..111J}}. The Fourier expansion of the magnitude in terms of the phase is given as \citep{deb2009}:
\begin{equation}\label{eq:fourier}
    m(\Phi) = m_0 +  \sum_{i=1}^{n} A_i \sin [2\pi i(\Phi) + \phi_i], 
\end{equation}

where $m(\Phi)$ is the extinction-corrected apparent magnitude in $G$ band, $m_0$ is the mean magnitude, $n$ is the order of expansion, $\Phi$ is the phase obtained from the phased light curve varying from 0 to 1. $A_i$ and $\phi_i$ are the i-th order Fourier amplitude and phase coefficients, respectively.

We determine the Fourier parameters by locally fitting Eq. \ref{eq:fourier} to the phased light curve. We use the \textit{curve\_fit} routine of the \texttt{SciPy}\footnote{\url{https://docs.scipy.org/doc/scipy/reference/generated/scipy.optimize.curve_fit.html}} module, which uses the Levenberg-Marquardt (LM) algorithm, widely used in the case of non-linear least squares fit. We found  that a fourth-order (n = 4) fit was sufficient to reproduce the shape of the light curve. Any order greater than n = 4, would overfit the curve. We also compared our independently extracted Fourier phase differences ($\phi_{31, \rm TW}$) with the corresponding values from the \textit{Gaia} DR3 automated pipeline ($\phi_{31, \rm Gaia}$). As illustrated in the right panel of Figure \ref{fig:comparison_gaia}, the values follow the 1:1 relation but exhibit a small scatter of 0.13 rad and a minor bias of 0.03 rad. Although both approaches utilize Levenberg-Marquardt non-linear fitting \citet{2023AA...674A..18C}, these subtle discrepancies are common when comparing a manually curated, sigma-clipped sample against a fully automated, survey-level pipeline. The variations primarily stem from differences in light curve pre-processing (outlier rejection) and the exact number of Fourier harmonics utilized in the fit model. The impact of utilizing our independent Fourier parameters versus the literature values is discussed in further detail in Section \ref{subsec:lit}

\citet{1981ApJ...248..291S} first demonstrated that a certain combination of these Fourier parameters was directly related to some physical parameters of the pulsating star. These are defined either as a linear combination of the parameters or the ratios of Fourier amplitudes, e.g.,
 \begin{equation}\label{eq:linear}
     \phi_{i1} = \phi_i - i\phi_1\,,
 \end{equation}
 \begin{equation}\label{eq:ratio}
     R_{i1} = \frac{A_i}{A_1}\,.
 \end{equation}

\begin{table*}[]
    \caption{The calibration dataset of 72 RRLs from Gaia~DR3 along with the spectroscopic [Fe/H] and their errors as obtained from \citet{2021MNRAS.503.4719G} }
    \scalebox{0.9}{
    \begin{tabular}{lcccc}
    \hline
        Gaia DR3 Source ID & $l$ & $b$ & $A_G$ & [Fe/H]$_{\rm spec}$ (dex)  \\ \hline      
        6226585956422527616 & 337.4 & 27.6 & 0.086 & -1.61 $\pm$ 0.22 \\ 
        630421935431871232 & 208.4 & 53.1 & 0.039 & -1.49 $\pm$ 0.10 \\ 
        6380659528686603008 & 311.2 & -43.2 & 0.019 & -1.79 $\pm$ 0.10 \\ 
        6483680332235888896 & 355.3 & -43.1 & 0.012 & -1.60 $\pm$ 0.14 \\ 
        6570585628216929408 & 1.0 & -55.6 & 0.007 & -2.13 $\pm$ 0.10 \\ 
        6573170751851975936 & 0.6 & -54.2 & 0.008 & -2.89 $\pm$ 0.10 \\ 
        6701724002113004928 & 340.6 & -14.5 & 0.094 & -1.70 $\pm$ 0.13 \\ 
        6730211038418525056 & 358.1 & -15.6 & 0.058 & -1.55 $\pm$ 0.13 \\ 
        6787617919184986496 & 15.7 & -43.2 & 0.079 & -1.65 $\pm$ 0.13 \\ 
        1009665142487836032 & 176.1 & 41.7 & 0.014 & -1.64 $\pm$ 0.10 \\ 
    \hline
    \end{tabular}}
    \begin{tablenotes}
      \small
      \item Note: Full table is available in machine-readable format as supplementary material. 
    \end{tablenotes}
    \label{tab:extinction}
\end{table*}

\begin{figure}
\centering
\includegraphics[width=0.8\linewidth]{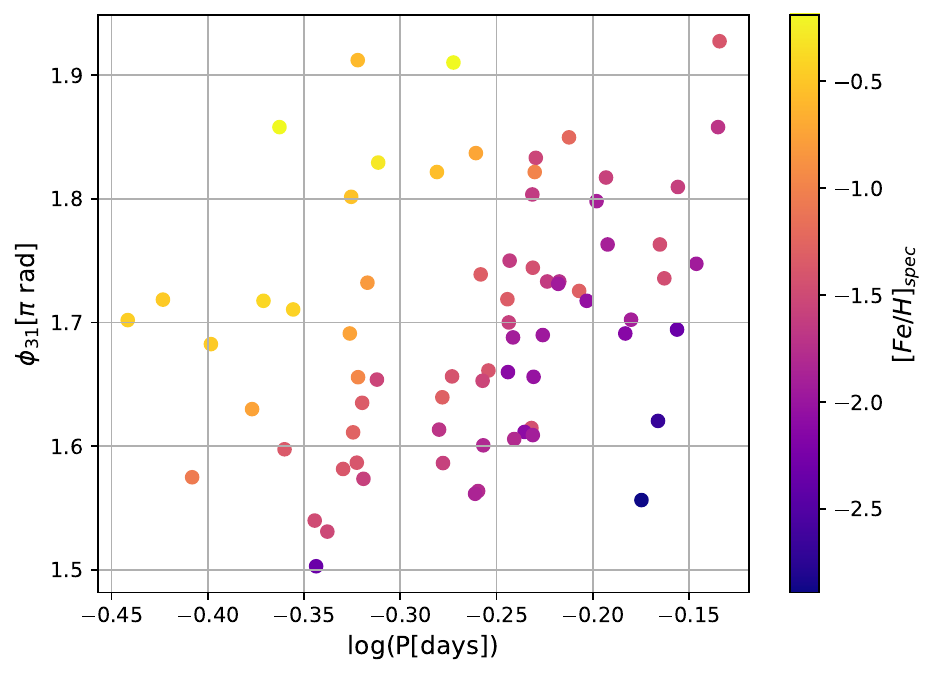}
\caption[]{Gaia $G$-band $P$-$\phi_{31}$-[Fe/H]. Period vs. $\phi_{31}$ plane with stars color-coded based on the metallicity adopted from \citet{2021MNRAS.503.4719G} }
\label{fig001}
\end{figure}

 We utilized $\phi_{31}$ for our analysis because its relation to the period offers the clearest separation of metallicities with the least scatter (Fig. \ref{fig001}; \citet{1981ApJ...248..291S, 1996AA...312..111J}). This traditional feature selection is strongly supported by recent machine-learning analyses of \textit{Gaia} DR3 RRLs; notably, \citet{2025MNRAS.536.2749M} concluded that $\phi_{31}$ remains the most influential parameter for predicting RRab metallicity, and that incorporating additional, higher-order Fourier parameters risks introducing unnecessary noise rather than improving the predictive accuracy. Following \citet{1996AA...312..111J}, we corrected for the 2$\pi$-periodic ambiguity by shifting phases closer to the mean. To ensure the reliability of the derived physical parameters, we applied an empirical data-quality cut to remove outliers with poorly constrained Fourier fits ($\sigma_{\phi_{31}} > 0.3$ rad). This specific threshold was chosen to optimize the trade-off between minimizing the formal uncertainties of the derived Fourier parameters and maintaining a statistically robust sample size. After applying this cut, 72 stars remained for the final calibration (see Table \ref{tab:extinction}).

The formal uncertainties for the individual Fourier coefficients ($\sigma_{\phi_i}$) were derived from the covariance matrix generated during the least-squares fitting process. The corresponding uncertainty for the derived phase parameter, $\phi_{i1}$, was then calculated using standard error propagation, assuming the covariances between the parameters are negligible \citep{2010MNRAS.402..691D}:
\begin{equation}
\sigma_{\phi_{i1}} = \sqrt{\sigma_{\phi_i}^2 + i^2\sigma_{\phi_1}^2}.
\end{equation}

\subsection{The $P-\phi_{31}-\rm{[Fe/H]}$ Relation}\label{sec:analysis}

We have fitted a linear relation between [Fe/H], period and $\phi_{31}$ of the form \citep{1996AA...312..111J, article}:

\begin{equation}\label{eq:fit}
    \mathrm{[Fe/H]} = a + b(P) + c(\phi_{31})\,.
\end{equation}
 We adopted the Orthogonal Distance Regression (ODR) routine which is a part of the \texttt{SciPy}\footnote{\url{https://docs.scipy.org/doc/scipy/reference/odr.html}} module for multiple-variable linear regression. This method utilizes a modified trust-region LM-type algorithm to estimate the best fitting parameters. The primary reason to choose ODR for fitting is because it minimizes the perpendicular distance to the fit with no differentiation between dependent and independent variables. This is necessary because metallicity here is estimated indirectly, and its measurements might have inherent uncertainties. Also, it minimizes the sum of the squared perpendicular distances between the data points and the fitted model. This approach considers both horizontal and vertical deviations, ensuring a good fit not only along the x-axis (period) but also along the y-axis ($\phi_{31}$). The final predicted photometric metallicities ($\rm[Fe/H]_{TW}$) evaluated from this regression are cataloged alongside the input parameters for each star in Table \ref{tab:final}.

 \begin{table*}[]
    \centering
    \caption{Table consists of the Gaia ID of stars in our calibration sample along with their derived periods, Fourier parameters ($\phi_{31}$ and $R_{21}$) and [Fe/H] predicted by our work ($\rm{[Fe/H]_{TW}}$ Eq. \ref{eq:final}) and the respective errors in each of the parameters.}
    \scalebox{0.9}{
    \begin{tabular}{lcccccc}
    \hline
        Gaia DR3 Source ID & Period (days) & $\phi_{31}$ (rad) & $R_{21}$  & [Fe/H]$_{\rm TW}$(dex)  \\ \hline
        6226585956422527616 & 0.6985 & 5.69 $\pm$ 0.12 & 0.49 $\pm$ 0.03 & -1.75 $\pm$ 0.88 \\ 
        630421935431871232 & 0.4524 & 4.84 $\pm$ 0.10 & 0.43 $\pm$ 0.02 & -1.66 $\pm$ 0.79 \\ 
        6380659528686603008 & 0.5501 & 4.91 $\pm$ 0.18 & 0.48 $\pm$ 0.04 & -2.13 $\pm$ 0.84 \\ 
        6483680332235888896 & 0.4796 & 4.94 $\pm$ 0.21 & 0.43 $\pm$ 0.06 & -1.65 $\pm$ 0.86 \\ 
        6570585628216929408 & 0.5700 & 5.22 $\pm$ 0.18 & 0.43 $\pm$ 0.04 & -1.75 $\pm$ 0.86 \\ 
        6573170751851975936 & 0.6687 & 4.89 $\pm$ 0.11 & 0.35 $\pm$ 0.03 & -2.88 $\pm$ 0.82 \\ 
        6701724002113004928 & 0.5250 & 5.07 $\pm$ 0.09 & 0.42 $\pm$ 0.02 & -1.72 $\pm$ 0.81 \\ 
        6730211038418525056 & 0.5893 & 5.76 $\pm$ 0.11 & 0.46 $\pm$ 0.03 & -0.96 $\pm$ 0.87 \\ 
        6787617919184986496 & 0.5869 & 5.67 $\pm$ 0.06 & 0.45 $\pm$ 0.01 & -1.10 $\pm$ 0.85 \\ 
        1009665142487836032 & 0.5974 & 5.45 $\pm$ 0.07 & 0.48 $\pm$ 0.02 & -1.53 $\pm$ 0.84 \\  
    \hline
    \end{tabular}
    }
    \label{tab:final}
    \begin{tablenotes}
      \small
      \item Note: Full table is available in machine-readable format as supplementary material.
    \end{tablenotes}
\end{table*}

Our best fit $G$-band $P$-$\phi_{31}$-[Fe/H] relation based on the Gaia DR3 light curves of 72 finally selected RRab stars is:
\begin{equation}\label{eq:final}
\begin{aligned}
       \relax \mathrm{[Fe/H]} = -6.93 \pm 0.58 - (6.04 \pm 0.37)P + (1.65 \pm 0.11)\phi_{31} ; \sigma = 0.26\,
\end{aligned}
\end{equation}
To determine the theoretical uncertainties in our predicted photometric metallicities, we rigorously propagated the formal errors of the ODR fit parameters along with the intrinsic measurement uncertainties of the independent variables. Because the ODR algorithm fits a multi-dimensional plane, the fitted coefficients are inherently correlated. Therefore, we utilized the full covariance matrix ($C_{ij}$) from the ODR output to properly account for parameter anti-correlation. The total theoretical uncertainty for a given star is calculated as:
\begin{equation}\label{eq:error_prop}
\sigma_{\mathrm{[Fe/H]}}^2 = C_{00} + P^2 C_{11} + \phi_{31}^2 C_{22} + 2P C_{01} + 2\phi_{31} C_{02} + 2P\phi_{31} C_{12} + \beta_1^2 \sigma_P^2 + \beta_2^2 \sigma_{\phi_{31}}^2
\end{equation}
where $\beta_1$ and $\beta_2$ are the fitted coefficients for $P$ and $\phi_{31}$, respectively, and $\sigma_P$ and $\sigma_{\phi_{31}}$ are the measurement errors. Using this full covariance formulation, the average theoretical uncertainty of our sample is equal to 0.26 dex. The calibration of our relation is done with the spectroscopic metallicities of \citet{2021MNRAS.503.4719G}. The empirical RMS dispersion in our relation with respect to the calibration data is 0.26 dex, which is in excellent statistical agreement with our theoretical error propagation.


\section{Discussion}\label{sec:discussion}
\subsection{Comparison with spectroscopic metallicities}\label{subsec:spec}
\begin{figure*}
\centering
\includegraphics[scale=0.8]{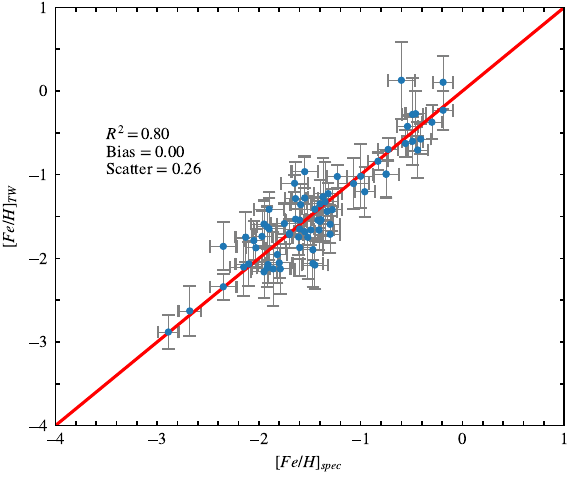}
\caption[]{Comparison of [Fe/H] predicted by our relation with that of the respective spectroscopic [Fe/H]. The error bars for the predicted photometric metallicities are plotted with the relative errors of [Fe/H] obtained considering errors in the fitting parameters a, b and c, and the Fourier parameter $\phi_{31}$. The red solid line represents the $R^2 = 1$ relation.}
\label{fig000}
\end{figure*}
In order to test our $G$-band relation, we take metallicities within the calibration range $-2.88 \leq \rm{[Fe/H]} \leq 0.13$ (dex). We have plotted the spectroscopic [Fe/H] of these stars with the corresponding [Fe/H] predicted using their respective $\phi_{31}$ and period from Gaia DR3 $G$-band light curves. In Fig. \ref{fig000}, we compare the spectroscopic [Fe/H] of these stars with their corresponding estimated photometric [Fe/H] from this work. The error bars are plotted to show the relative error in the [Fe/H] predicted by our relation. 
\begin{figure*}
\centering
\includegraphics{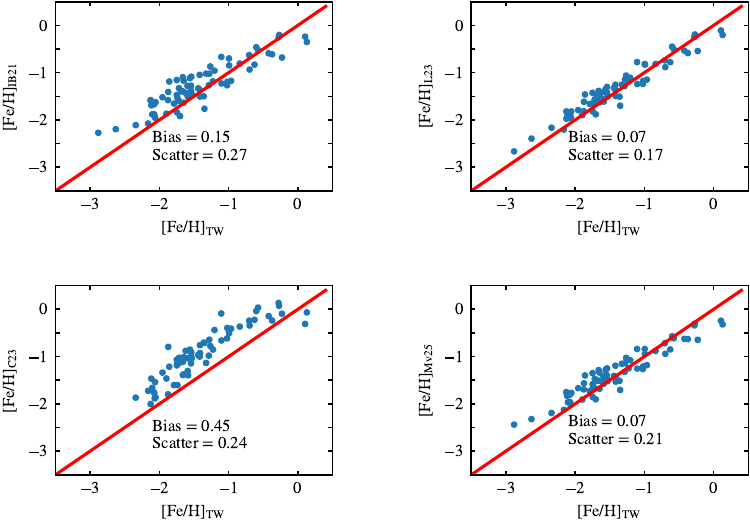}
\caption[]{Comparison of the [Fe/H] derived with our $G$-band $P$-$\phi_{31}$-[Fe/H] relation with that derived using the relationship of \citetalias{2021MNRAS.502.5686I} (top left), \citetalias{2023AA...674A..18C} (bottom left), \citetalias{2023ApJ...944...88L} (top right) and \citetalias{2025MNRAS.536.2749M} (bottom right), applied to the $\phi_{31}$ values derived in the Gaia DR3 dataset. The red solid line represents the $R^2=1$ relation. Each plot has its bias and scatter values mentioned separately.}
\label{fig002}
\end{figure*}
The regression model achieved $R^2 = 0.80$, indicating that it explains 80\% of the variance in the observed [Fe/H] values. This result suggests a good fit between the model predictions and the spectroscopic metallicities. Furthermore, the negligible bias of $0.00$ dex demonstrates excellent alignment with the spectroscopic data, indicating that the model exhibits no measurable systematic offset from the true [Fe/H] values. Finally, a scatter ($\sigma$) of $0.26$ dex suggests that the predictions are reasonably clustered around the observed values, with deviations of up to $0.26$ dex in either direction.

These combined findings provide strong evidence that our $G$-band metallicity relation is robust for estimating [Fe/H] in RRab stars.

\subsection{Comparison with literature}\label{subsec:lit}

We compared our $G$-band $P$-$\phi_{31}$-[Fe/H] relation (Eq. \ref{eq:final}) with previous relations found in the literature for similar passbands. In particular, we focus on the relations found in \citet{2021MNRAS.502.5686I} (hereafter IB21), \citet{2023AA...674A..18C} (hereafter C23), \citet{2023ApJ...944...88L} (hereafter L23) and \citet{2025MNRAS.536.2749M} (hereafter Mv25) in Fig. \ref{fig002}. 

\citetalias{2021MNRAS.502.5686I} introduced a $G$-band $P$-$\phi_{31}$-[Fe/H] relationship for RRab stars which were calibrated using a sample of 84 stars found by cross-matching the spectroscopic sample of \citet{1994AJ....108.1016L} with the SOS (Specific Object Study; \citealt{2019AA...622A..60C}) RRL catalogue which is based on Gaia DR2 light curves. To set the coefficients on the same scale, an additional offset of $\pi$ rad is subtracted from $\phi_{31}$ as the SOS $\phi_{31}$ are reported in the Kepler photometric scale. Their $P$-$\phi_{31}$-[Fe/H] relation reads as:
\begin{equation} \label{eq:ib20}
\begin{aligned}
    \relax     \mathrm{[Fe/H]_{IB21}} = -1.68 - 5.08(P - 0.6) + 0.68(\phi_{31} - 2.0) 
\end{aligned}
\end{equation}
Metallicity in  \citetalias{2021MNRAS.502.5686I} was on the scale of \citet{1984ApJS...55...45Z}(ZW84) and thus was converted using the following relation given by \citet{2009AA...508..695C}(C09) :
\begin{equation}
\begin{aligned}
    \relax \mathrm{[Fe/H]_{C09}} = 1.105\mathrm{[Fe/H]_{ZW84}} + 0.160
\end{aligned}
\end{equation}

\citetalias{2023AA...674A..18C} does not have an explicit relationship but their catalogue (\texttt{vari\_rrlyrae}) contains the photometric [Fe/H] determined by the SOS Ceph\&RRL pipeline for the Gaia DR3 light curves. We cross-matched the catalogue with ours to find 69 common RRab sources. We compare the photometric [Fe/H] derived from our relation to that of \texttt{vari\_rrlyrae}.

\citetalias{2023ApJ...944...88L} built a linear $P$-$\phi_{31}$-$R_{21}$-[Fe/H] relationship inspired by the comprehensive analysis of \citet{2021ApJ...920...33D}. They cross-matched the spectroscopic sample of \citet{2022MNRAS.517.2787L} with the photometric sample of \citetalias{2023AA...674A..18C} and derived the following relation for 2046 RRab sources:

\begin{equation}\label{eq:L23}
\begin{aligned}
        \relax \mathrm{[Fe/H]_{L23}} = -1.888 \pm 0.002 - (5.772 \pm 0.026)(P - 0.6) + (1.090 \pm 0.005)(\phi_{31} - 2) \\
        + (1.065 \pm 0.030)(R_{21} - 0.45) 
\end{aligned}
\end{equation}

To check if adding an extra $R_{21}$ term to our relation improves the fit, we obtained a plot  of photometric metallicities derived from Eq. \ref{eq:final} versus the one derived from the following equation:

\begin{equation}\label{eq:r21}
\begin{aligned}
        \relax \mathrm{[Fe/H]} = -7.55 \pm 0.63 - (5.81 \pm 0.37)P + (1.57 \pm 0.11)(\phi_{31} - 2) + (1.96 \pm 0.68)R_{21} 
\end{aligned}
\end{equation}

\begin{figure}
\centering
\includegraphics[scale=0.8]{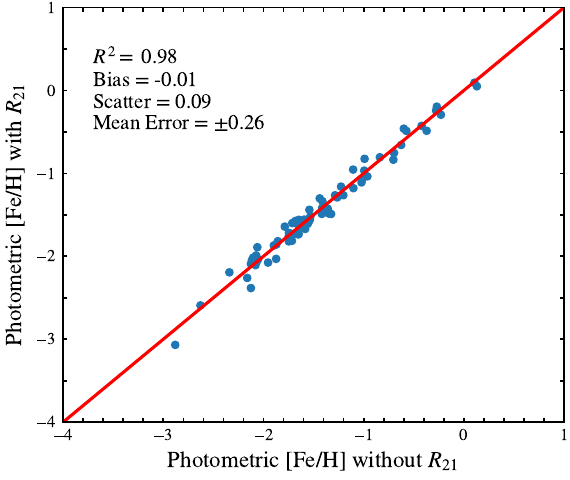}
\caption[]{Comparison of photometric metallicities obtained using Eq. \ref{eq:final} and Eq. \ref{eq:r21}}
\label{fig:comp}
\end{figure}

Figure \ref{fig:comp} shows that there is no significant difference in the photometric [Fe/H] with an additional $R_{21}$ term. While the inclusion of the $R_{21}$ parameter does not yield a statistically significant improvement in the overall metallicity estimates, we note that the correlation coefficient between the predictions of the two models remains nearly 1 ($R \approx 1.0$). Furthermore, our methodology and derived offsets can be contextualized alongside the recent work by \citet{2023MNRAS.525.3486J}. Utilizing Gaia DR3 data, they demonstrated that the raw photometric metallicities for RR Lyrae stars can be highly unreliable and require revised calibrations. This aligns with our findings regarding the necessity of independent calibrations and careful parameter selection to obtain robust metallicities. 

We also compared our relation with that of \citetalias{2025MNRAS.536.2749M}. They utilized the $\phi_{31}$ and period values from \texttt{vari\_rrlyrae} table of \citetalias{2023AA...674A..18C} and used the Bayesian approach to derive their relation:
\begin{equation}\label{eq:Mv25}
\begin{aligned}
    \relax \mathrm{[Fe/H]_{Mv25}} = (-5.55 \pm 0.33)P + (0.94 \pm 0.09)\phi_{31} - (0.37 \pm 0.20)
\end{aligned}
\end{equation}

It is to be noted that \citetalias{2023AA...674A..18C} uses a Fourier expansion in cosine function so an additional $\pi$ offset had to be subtracted from $\phi_{31}$ while comparing our relation with that of \citetalias{2023ApJ...944...88L} and \citetalias{2025MNRAS.536.2749M}. 

From Fig. \ref{fig002} it is clear that our $G$-band $P$-$\phi_{31}$-[Fe/H] relation aligns most closely with those of \citetalias{2023ApJ...944...88L} and \citetalias{2025MNRAS.536.2749M}, both of which are based on Gaia DR3 light curves and SOS-derived Fourier parameters. The small positive offset seen relative to \citetalias{2025MNRAS.536.2749M} likely reflects differences in the construction of their regression model rather than any intrinsic discrepancy in the underlying Fourier quantities. Although \citetalias{2023AA...674A..18C} also utilizes DR3 photometry, their metallicities originate from the non-linear SOS calibration inspired by \citet{2013ApJ...773..181N}, leading naturally to a systematic overestimation when compared to our linear relation. To quantify this, the comparison for the 69 common stars shows a mean offset of only $0.03$ rad (with a scatter of $0.13$ rad) between our $\phi_{31}$ values and the SOS $\phi_{31}$ values (Fig. \ref{fig:comparison_gaia} right panel). Because this phase difference is so small, it propagates to a metallicity shift of only $\sim 0.04$ dex. Therefore, we conclude that the systematic metallicity offset observed does not primarily arise from differences in how the Fourier parameters were derived. Finally, the comparison with \citetalias{2021MNRAS.502.5686I} highlights the improvement in Gaia photometry from DR2 to DR3, as indicated by the reduced scatter and smaller bias.

While recent large-scale studies (e.g., \citet{2023AA...674A..18C, 2025MNRAS.536.2749M}) have similarly leveraged \textit{Gaia} DR3 photometry to derive [Fe/H] relations, our work distinguishes itself through its specific calibration methodology. The primary similarity lies in the foundational use of the $P$-$\phi_{31}$ framework and the native \textit{Gaia} $G$-band time-series. However, the critical difference is our approach to the calibration sample. Whereas large-scale surveys often rely on massive but heterogeneous calibration samples (incorporating metallicities derived from varying spectroscopic resolutions and pipelines), our relation is anchored exclusively to a highly curated, homogeneously analyzed sample of high-resolution spectroscopic metallicities from \citet{2021MNRAS.503.4719G}. 

The novelty of our work lies in prioritizing calibration purity and mathematical rigor over sample size. By restricting our fit to this high-fidelity spectroscopic sample and utilizing an Orthogonal Distance Regression (ODR) routine that fully accounts for parameter covariance, our derived relation minimizes the systematic biases and zero-point offsets that can plague heterogeneous datasets. Consequently, our calibration provides an independent, mathematically strict empirical baseline for $G$-band [Fe/H] estimation that complements the broader, machine-learning-driven relations currently in the literature.

\section{Summary and conclusions}\label{sec:conclusion}

In this work, we have presented a new, empirically calibrated $G$-band metallicity relationship for fundamental-mode RRLs (RRab) using data from the Gaia Data Release 3 (DR3). By combining high-quality photometry with robust spectroscopic metallicities, we derived a linear relation linking the pulsation period ($P$), the Fourier parameter $\phi_{31}$, and metallicity ([Fe/H]).

Our analysis was based on a carefully selected calibration sample of 72 RRab stars. While recent literature often employs massive, structurally heterogeneous datasets to derive photometric metallicities, our work establishes a complementary, high-purity empirical baseline. These targets were identified through visual inspection of light curves to ensure phase coverage and quality, and their metallicities were sourced exclusively from a homogeneously analyzed sample of high-resolution spectroscopic observations available in the literature. By restricting our fit to this highly curated sample, we minimize the systematic zero-point offsets that often affect results from large-scale surveys. The resulting $P$-$\phi_{31}$-[Fe/H] relation yields a high correlation coefficient ($R^2 = 0.80$), confirming the strength of the correlation between the photometric parameters and metallicity.

We validated the accuracy of our relation by comparing the predicted metallicities against the spectroscopic values of the calibration sample. By utilizing an Orthogonal Distance Regression (ODR) fitting routine and rigorously propagating theoretical uncertainties via the full covariance matrix, we ensured that our calibration is mathematically robust. The residuals indicate a negligible bias of $0.00$ dex and an empirical RMS scatter of 0.26 dex, which is in excellent statistical agreement with our theoretically derived errors. This demonstrates that the relation provides precise metallicity estimates for stars within the calibration range of $-2.88 \leq \text{[Fe/H]} \leq 0.13$ (dex). Furthermore, we investigated the inclusion of the amplitude ratio $R_{21}$ as an additional term in the regression model. Our tests revealed that this added complexity did not yield a statistically significant improvement in the metallicity estimation, justifying our adoption of the simpler period-$\phi_{31}$ formulation. Although the addition of this parameter did not significantly alter the predicted [Fe/H] values—yielding a correlation coefficient of nearly 1 between the two models—it ultimately introduced unnecessary mathematical complexity without improving predictive accuracy.

Finally, we compared our results with existing relations in the literature. We found:
\begin{itemize}
    \item Excellent agreement with the recent relations of \citetalias{2023ApJ...944...88L} and \citetalias{2025MNRAS.536.2749M}, which also utilize Gaia DR3 photometry. This consensus reinforces the reliability of the Gaia $G$-band photometry for metallicity determination.
    \item An improvement in precision (lower scatter) compared to the relation of \citetalias{2021MNRAS.502.5686I}, highlighting the superior quality of DR3 data over the preceding DR2 release.
    \item A systematic offset when compared to \citetalias{2023AA...674A..18C}, which we attribute to their use of a non-linear calibration approach versus the linear regression model employed in this work and other recent studies.
\end{itemize}

While our focus on calibration purity yields a robust and mathematically strict correlation, we caution that these results are inherently constrained by the relatively small size of our calibration sample (72 stars) and its limited metallicity coverage. Extrapolating these relations to RR Lyrae stars with metallicities significantly outside the bounds of our current sample should be approached with caution, as non-linearities may emerge in more extreme metallicity regimes. 

In conclusion, the updated $G$-band relation presented here offers a robust and straightforward tool for estimating the metallicities of RRLs. It is particularly well-suited for large-scale studies of galactic structure and stellar populations in the era of Gaia.






\authorcontributions{Conceptualization, H.P.S.; methodology, M, S.D., H.P.S.; software, M.; validation, M, S.D., H.P.S.; formal analysis, M.; investigation, M.; resources, M.; data curation, M, S.D., H.P.S.; writing---original draft preparation, M.; writing---review and editing, M, S.D., H.P.S., N.K.; visualization, M, S.D., H.P.S.; supervision, S.D., H.P.S.; project administration, H.P.S.; funding acquisition, H.P.S. All authors have read and agreed to the published version of the manuscript.}


\dataavailability{This work has made use of data from the European Space Agency (ESA) mission {\it Gaia} (\url{https://www.cosmos.esa.int/gaia}), processed by the {\it Gaia} Data Processing and Analysis Consortium (DPAC, \url{https://www.cosmos.esa.int/web/gaia/dpac/consortium}). Funding for the DPAC has been provided by national institutions, in particular the institutions participating in the {\it Gaia} Multilateral Agreement.}


\acknowledgments{The authors are grateful to the referees for useful suggestions that improved the quality of the manuscript. This work was completed as part of a Master’s thesis by the first author at the University of Delhi.}

\conflictsofinterest{The authors declare no conflicts of interest.}

\begin{adjustwidth}{-\extralength}{0cm}

\reftitle{References}


\PublishersNote{}
\end{adjustwidth}

\begin{thebibliography}{999}

\bibitem[{Catelan} et~al.(2004){Catelan}, {Pritzl}, and {Smith}]{2004ApJS..154..633C}
{Catelan}, M.; {Pritzl}, B.J.; {Smith}, H.A.
\newblock {The RR Lyrae Period-Luminosity Relation. I. Theoretical Calibration}.
\newblock {\em \apjs} {\bf 2004}, {\em 154},~633--649,  \href{http://arxiv.org/abs/astro-ph/0406067}{{\normalfont [arXiv:astro-ph/astro-ph/0406067]}}.
\newblock {\url{https://doi.org/10.1086/422916}}.

\bibitem[{Bono} et~al.(2001){Bono}, {Caputo}, {Castellani}, {Marconi}, and {Storm}]{2001MNRAS.326.1183B}
{Bono}, G.; {Caputo}, F.; {Castellani}, V.; {Marconi}, M.; {Storm}, J.
\newblock {Theoretical insights into the RR Lyrae K-band period-luminosity relation}.
\newblock {\em \mnras} {\bf 2001}, {\em 326},~1183--1190,  \href{http://arxiv.org/abs/astro-ph/0105481}{{\normalfont [arXiv:astro-ph/astro-ph/0105481]}}.
\newblock {\url{https://doi.org/10.1046/j.1365-8711.2001.04655.x}}.

\bibitem[{Longmore} et~al.(1985){Longmore}, {Fernley}, {Jameson}, {Sherrington}, and {Frank}]{1985MNRAS.216..873L}
{Longmore}, A.J.; {Fernley}, J.A.; {Jameson}, R.F.; {Sherrington}, M.R.; {Frank}, J.
\newblock {VJHK observations of the RR Lyrae star VY Serpentis.}
\newblock {\em \mnras} {\bf 1985}, {\em 216},~873--882.
\newblock {\url{https://doi.org/10.1093/mnras/216.4.873}}.

\bibitem[{Sollima} et~al.(2006){Sollima}, {Cacciari}, and {Valenti}]{2006MNRAS.372.1675S}
{Sollima}, A.; {Cacciari}, C.; {Valenti}, E.
\newblock {The RR Lyrae period-K-luminosity relation for globular clusters: an observational approach}.
\newblock {\em \mnras} {\bf 2006}, {\em 372},~1675--1680,  \href{http://arxiv.org/abs/astro-ph/0608397}{{\normalfont [arXiv:astro-ph/astro-ph/0608397]}}.
\newblock {\url{https://doi.org/10.1111/j.1365-2966.2006.10962.x}}.

\bibitem[{Braga} et~al.(2015){Braga}, {Dall'Ora}, {Bono}, {Stetson}, {Ferraro}, {Iannicola}, {Marengo}, {Neeley}, {Persson}, {Buonanno}, {Coppola}, {Freedman}, {Madore}, {Marconi}, {Matsunaga}, {Monson}, {Rich}, {Scowcroft}, and {Seibert}]{2015ApJ...799..165B}
{Braga}, V.F.; {Dall'Ora}, M.; {Bono}, G.; {Stetson}, P.B.; {Ferraro}, I.; {Iannicola}, G.; {Marengo}, M.; {Neeley}, J.; {Persson}, S.E.; {Buonanno}, R.;  et~al.
\newblock {On the Distance of the Globular Cluster M4 (NGC 6121) Using RR Lyrae Stars. I. Optical and Near-infrared Period-Luminosity and Period-Wesenheit Relations}.
\newblock {\em \apj} {\bf 2015}, {\em 799},~165,  \href{http://arxiv.org/abs/1411.6826}{{\normalfont [arXiv:astro-ph.GA/1411.6826]}}.
\newblock {\url{https://doi.org/10.1088/0004-637X/799/2/165}}.

\bibitem[{Muraveva} et~al.(2015){Muraveva}, {Palmer}, {Clementini}, {Luri}, {Cioni}, {Moretti}, {Marconi}, {Ripepi}, and {Rubele}]{2015ApJ...807..127M}
{Muraveva}, T.; {Palmer}, M.; {Clementini}, G.; {Luri}, X.; {Cioni}, M.R.L.; {Moretti}, M.I.; {Marconi}, M.; {Ripepi}, V.; {Rubele}, S.
\newblock {New Near-infrared Period-Luminosity-Metallicity Relations for RR Lyrae Stars and the Outlook for Gaia}.
\newblock {\em \apj} {\bf 2015}, {\em 807},~127,  \href{http://arxiv.org/abs/1505.06001}{{\normalfont [arXiv:astro-ph.SR/1505.06001]}}.
\newblock {\url{https://doi.org/10.1088/0004-637X/807/2/127}}.

\bibitem[Beaton et~al.(2018)Beaton, Bono, Braga, Dall’Ora, Fiorentino, Jang, Martínez-Vázquez, Matsunaga, Monelli, Neeley, and Salaris]{Beaton_2018}
Beaton, R.L.; Bono, G.; Braga, V.F.; Dall’Ora, M.; Fiorentino, G.; Jang, I.S.; Martínez-Vázquez, C.E.; Matsunaga, N.; Monelli, M.; Neeley, J.R.;  et~al.
\newblock Old-Aged Primary Distance Indicators.
\newblock {\em Space Science Reviews} {\bf 2018}, {\em 214}.
\newblock {\url{https://doi.org/10.1007/s11214-018-0542-1}}.

\bibitem[{Bhardwaj} et~al.(2020){Bhardwaj}, {Rejkuba}, {de Grijs}, {Herczeg}, {Singh}, {Kanbur}, and {Ngeow}]{2020AJ....160..220B}
{Bhardwaj}, A.; {Rejkuba}, M.; {de Grijs}, R.; {Herczeg}, G.J.; {Singh}, H.P.; {Kanbur}, S.; {Ngeow}, C.C.
\newblock {Near-infrared Census of RR Lyrae Variables in the Messier 3 Globular Cluster and the Period-Luminosity Relations}.
\newblock {\em \aj} {\bf 2020}, {\em 160},~220,  \href{http://arxiv.org/abs/2008.11745}{{\normalfont [arXiv:astro-ph.SR/2008.11745]}}.
\newblock {\url{https://doi.org/10.3847/1538-3881/abb3f9}}.

\bibitem[Zhang et~al.(2020)Zhang, Liu, Li, Deng, Yan, and Shi]{Zhang_2020}
Zhang, B.; Liu, C.; Li, C.Q.; Deng, L.C.; Yan, T.S.; Shi, J.R.
\newblock Exploring the spectral information content in the LAMOST medium-resolution survey (MRS).
\newblock {\em Research in Astronomy and Astrophysics} {\bf 2020}, {\em 20},~051.
\newblock {\url{https://doi.org/10.1088/1674-4527/20/4/51}}.

\bibitem[{Cui} et~al.(2012){Cui}, {Zhao}, {Chu}, {Li}, {Li}, {Zhang}, {Su}, {Yao}, {Wang}, {Xing}, {Li}, {Zhu}, {Wang}, {Gu}, {Luo}, {Xu}, {Zhang}, {Liu}, {Zhang}, {Yang}, {Cao}, {Chen}, {Chen}, {Chen}, {Chen}, {Chu}, {Feng}, {Gong}, {Hou}, {Hu}, {Hu}, {Hu}, {Jia}, {Jiang}, {Jiang}, {Jiang}, {Jin}, {Li}, {Li}, {Li}, {Liu}, {Liu}, {Lu}, {Mao}, {Men}, {Qi}, {Qi}, {Shi}, {Tang}, {Tao}, {Wang}, {Wang}, {Wang}, {Wang}, {Wang}, {Wang}, {Wang}, {Wang}, {Wang}, {Wang}, {Wang}, {Wang}, {Xu}, {Xu}, {Yang}, {Yu}, {Yuan}, {Yuan}, {Zhai}, {Zhang}, {Zhang}, {Zhang}, {Zhao}, {Zhou}, {Zhou}, {Zhu}, and {Zou}]{2012RAA....12.1197C}
{Cui}, X.Q.; {Zhao}, Y.H.; {Chu}, Y.Q.; {Li}, G.P.; {Li}, Q.; {Zhang}, L.P.; {Su}, H.J.; {Yao}, Z.Q.; {Wang}, Y.N.; {Xing}, X.Z.;  et~al.
\newblock {The Large Sky Area Multi-Object Fiber Spectroscopic Telescope (LAMOST)}.
\newblock {\em Research in Astronomy and Astrophysics} {\bf 2012}, {\em 12},~1197--1242.

\bibitem[{Ferreira Lopes} et~al.(2025){Ferreira Lopes}, {Guti{\'e}rrez-Soto}, {Ferreira Alberice}, {Monsalves}, {Hazarika}, {Catelan}, {Placco}, {Limberg}, {Almeida-Fernandes}, {Perottoni}, {Smith Castelli}, {Akras}, {Alonso-Garc{\'\i}a}, {Cordeiro}, {Jaque Arancibia}, {Daflon}, {Dias}, {Gon{\c{c}}alves}, {Machado-Pereira}, {Lopes}, {Bom}, {Thom de Souza}, {de Is{\'\i}dio}, {Alvarez-Candal}, {De Rossi}, {Bonatto}, {Cubillos Palma}, {Borges Fernandes}, {Humire}, {Oliveira Schwarz}, {Schoenell}, {Kanaan}, and {Mendes de Oliveira}]{2025AA...693A.306F}
{Ferreira Lopes}, C.E.; {Guti{\'e}rrez-Soto}, L.A.; {Ferreira Alberice}, V.S.; {Monsalves}, N.; {Hazarika}, D.; {Catelan}, M.; {Placco}, V.M.; {Limberg}, G.; {Almeida-Fernandes}, F.; {Perottoni}, H.D.;  et~al.
\newblock {Stellar atmospheric parameters and chemical abundances of \textasciitilde5 million stars from S-PLUS multiband photometry}.
\newblock {\em \aap} {\bf 2025}, {\em 693},~A306.

\bibitem[{Das} et~al.(2018){Das}, {Bhardwaj}, {Kanbur}, {Singh}, and {Marconi}]{2018MNRAS.481.2000D}
{Das}, S.; {Bhardwaj}, A.; {Kanbur}, S.M.; {Singh}, H.P.; {Marconi}, M.
\newblock {On the variation of light-curve parameters of RR Lyrae variables at multiple wavelengths}.
\newblock {\em \mnras} {\bf 2018}, {\em 481},~2000--2017,  \href{http://arxiv.org/abs/1808.08165}{{\normalfont [arXiv:astro-ph.SR/1808.08165]}}.
\newblock {\url{https://doi.org/10.1093/mnras/sty2358}}.

\bibitem[{Jurcsik} and {Kovacs}(1996)]{1996AA...312..111J}
{Jurcsik}, J.; {Kovacs}, G.
\newblock {Determination of [Fe/H] from the light curves of RR Lyrae stars.}
\newblock {\em \aap} {\bf 1996}, {\em 312},~111--120.

\bibitem[Smolec(2005)]{article}
Smolec, R.
\newblock Metallicity dependence of the Blazhko effect.
\newblock {\em \actaa} {\bf 2005}, {\em 55}.

\bibitem[{Nemec} et~al.(2013){Nemec}, {Cohen}, {Ripepi}, {Derekas}, {Moskalik}, {Sesar}, {Chadid}, and {Bruntt}]{2013ApJ...773..181N}
{Nemec}, J.M.; {Cohen}, J.G.; {Ripepi}, V.; {Derekas}, A.; {Moskalik}, P.; {Sesar}, B.; {Chadid}, M.; {Bruntt}, H.
\newblock {Metal Abundances, Radial Velocities, and Other Physical Characteristics for the RR Lyrae Stars in the Kepler Field}.
\newblock {\em \apj} {\bf 2013}, {\em 773},~181,  \href{http://arxiv.org/abs/1307.5820}{{\normalfont [arXiv:astro-ph.SR/1307.5820]}}.
\newblock {\url{https://doi.org/10.1088/0004-637X/773/2/181}}.

\bibitem[{Martinez-Vazquez} et~al.(2016){Martinez-Vazquez}, {Monelli}, {Bono}, {Stetson}, {Gallart}, {Bernard}, {Fiorentino}, and {Dall'Ora}]{2016CoKon.105...53M}
{Martinez-Vazquez}, C.E.; {Monelli}, M.; {Bono}, G.; {Stetson}, P.B.; {Gallart}, C.; {Bernard}, E.J.; {Fiorentino}, G.; {Dall'Ora}, M.
\newblock {A new Phi\_31-period-metallicity relation for RR Lyrae stars}.
\newblock {\em Communications of the Konkoly Observatory Hungary} {\bf 2016}, {\em 105},~53--56.

\bibitem[{Ngeow} et~al.(2016){Ngeow}, {Yu}, {Bellm}, {Yang}, {Chang}, {Miller}, {Laher}, {Surace}, and {Ip}]{2016ApJS..227...30N}
{Ngeow}, C.C.; {Yu}, P.C.; {Bellm}, E.; {Yang}, T.C.; {Chang}, C.K.; {Miller}, A.; {Laher}, R.; {Surace}, J.; {Ip}, W.H.
\newblock {The Palomar Transient Factory and RR Lyrae: The Metallicity-Light Curve Relation Based on ab-type RR Lyrae in the Kepler Field}.
\newblock {\em \apjs} {\bf 2016}, {\em 227},~30,  \href{http://arxiv.org/abs/1612.03366}{{\normalfont [arXiv:astro-ph.SR/1612.03366]}}.
\newblock {\url{https://doi.org/10.3847/1538-4365/227/2/30}}.

\bibitem[{Iorio} and {Belokurov}(2021)]{2021MNRAS.502.5686I}
{Iorio}, G.; {Belokurov}, V.
\newblock {Chemo-kinematics of the Gaia RR Lyrae: the halo and the disc}.
\newblock {\em \mnras} {\bf 2021}, {\em 502},~5686--5710,  \href{http://arxiv.org/abs/2008.02280}{{\normalfont [arXiv:astro-ph.GA/2008.02280]}}.
\newblock {\url{https://doi.org/10.1093/mnras/stab005}}.

\bibitem[{Mullen} et~al.(2021){Mullen}, {Marengo}, {Mart{\'\i}nez-V{\'a}zquez}, {Neeley}, {Bono}, {Dall'Ora}, {Chaboyer}, {Th{\'e}venin}, {Braga}, {Crestani}, {Fabrizio}, {Fiorentino}, {Gilligan}, {Monelli}, and {Stetson}]{2021ApJ...912..144M}
{Mullen}, J.P.; {Marengo}, M.; {Mart{\'\i}nez-V{\'a}zquez}, C.E.; {Neeley}, J.R.; {Bono}, G.; {Dall'Ora}, M.; {Chaboyer}, B.; {Th{\'e}venin}, F.; {Braga}, V.F.; {Crestani}, J.;  et~al.
\newblock {Metallicity of Galactic RR Lyrae from Optical and Infrared Light Curves. I. Period-Fourier-Metallicity Relations for Fundamental-mode RR Lyrae}.
\newblock {\em \apj} {\bf 2021}, {\em 912},~144,  \href{http://arxiv.org/abs/2103.09372}{{\normalfont [arXiv:astro-ph.SR/2103.09372]}}.
\newblock {\url{https://doi.org/10.3847/1538-4357/abefd4}}.

\bibitem[{Borucki} et~al.(2010){Borucki}, {Koch}, {Basri}, {Batalha}, {Brown}, {Caldwell}, {Caldwell}, {Christensen-Dalsgaard}, {Cochran}, {DeVore}, {Dunham}, {Dupree}, {Gautier}, {Geary}, {Gilliland}, {Gould}, {Howell}, {Jenkins}, {Kondo}, {Latham}, {Marcy}, {Meibom}, {Kjeldsen}, {Lissauer}, {Monet}, {Morrison}, {Sasselov}, {Tarter}, {Boss}, {Brownlee}, {Owen}, {Buzasi}, {Charbonneau}, {Doyle}, {Fortney}, {Ford}, {Holman}, {Seager}, {Steffen}, {Welsh}, {Rowe}, {Anderson}, {Buchhave}, {Ciardi}, {Walkowicz}, {Sherry}, {Horch}, {Isaacson}, {Everett}, {Fischer}, {Torres}, {Johnson}, {Endl}, {MacQueen}, {Bryson}, {Dotson}, {Haas}, {Kolodziejczak}, {Van Cleve}, {Chandrasekaran}, {Twicken}, {Quintana}, {Clarke}, {Allen}, {Li}, {Wu}, {Tenenbaum}, {Verner}, {Bruhweiler}, {Barnes}, and {Prsa}]{2010Sci...327..977B}
{Borucki}, W.J.; {Koch}, D.; {Basri}, G.; {Batalha}, N.; {Brown}, T.; {Caldwell}, D.; {Caldwell}, J.; {Christensen-Dalsgaard}, J.; {Cochran}, W.D.; {DeVore}, E.;  et~al.
\newblock {Kepler Planet-Detection Mission: Introduction and First Results}.
\newblock {\em Science} {\bf 2010}, {\em 327},~977.
\newblock {\url{https://doi.org/10.1126/science.1185402}}.

\bibitem[{Udalski} et~al.(2015){Udalski}, {Szyma{\'n}ski}, and {Szyma{\'n}ski}]{2015AcA....65....1U}
{Udalski}, A.; {Szyma{\'n}ski}, M.K.; {Szyma{\'n}ski}, G.
\newblock {OGLE-IV: Fourth Phase of the Optical Gravitational Lensing Experiment}.
\newblock {\em \actaa} {\bf 2015}, {\em 65},~1--38,  \href{http://arxiv.org/abs/1504.05966}{{\normalfont [arXiv:astro-ph.SR/1504.05966]}}.
\newblock {\url{https://doi.org/10.48550/arXiv.1504.05966}}.

\bibitem[{Law} et~al.(2009){Law}, {Kulkarni}, {Dekany}, {Ofek}, {Quimby}, {Nugent}, {Surace}, {Grillmair}, {Bloom}, {Kasliwal}, {Bildsten}, {Brown}, {Cenko}, {Ciardi}, {Croner}, {Djorgovski}, {van Eyken}, {Filippenko}, {Fox}, {Gal-Yam}, {Hale}, {Hamam}, {Helou}, {Henning}, {Howell}, {Jacobsen}, {Laher}, {Mattingly}, {McKenna}, {Pickles}, {Poznanski}, {Rahmer}, {Rau}, {Rosing}, {Shara}, {Smith}, {Starr}, {Sullivan}, {Velur}, {Walters}, and {Zolkower}]{2009PASP..121.1395L}
{Law}, N.M.; {Kulkarni}, S.R.; {Dekany}, R.G.; {Ofek}, E.O.; {Quimby}, R.M.; {Nugent}, P.E.; {Surace}, J.; {Grillmair}, C.C.; {Bloom}, J.S.; {Kasliwal}, M.M.;  et~al.
\newblock {The Palomar Transient Factory: System Overview, Performance, and First Results}.
\newblock {\em \pasp} {\bf 2009}, {\em 121},~1395,  \href{http://arxiv.org/abs/0906.5350}{{\normalfont [arXiv:astro-ph.IM/0906.5350]}}.
\newblock {\url{https://doi.org/10.1086/648598}}.

\bibitem[{Clementini} et~al.(2023){Clementini}, {Ripepi}, {Garofalo}, {Molinaro}, {Muraveva}, {Leccia}, {Rimoldini}, {Holl}, {Jevardat de Fombelle}, {Sartoretti}, {Marchal}, {Audard}, {Nienartowicz}, {Andrae}, {Marconi}, {Szabados}, {Evans}, {Lecoeur-Taibi}, {Mowlavi}, {Musella}, and {Eyer}]{2023AA...674A..18C}
{Clementini}, G.; {Ripepi}, V.; {Garofalo}, A.; {Molinaro}, R.; {Muraveva}, T.; {Leccia}, S.; {Rimoldini}, L.; {Holl}, B.; {Jevardat de Fombelle}, G.; {Sartoretti}, P.;  et~al.
\newblock {Gaia Data Release 3. Specific processing and validation of all-sky RR Lyrae and Cepheid stars: The RR Lyrae sample}.
\newblock {\em \aap} {\bf 2023}, {\em 674},~A18,  \href{http://arxiv.org/abs/2206.06278}{{\normalfont [arXiv:astro-ph.SR/2206.06278]}}.
\newblock {\url{https://doi.org/10.1051/0004-6361/202243964}}.

\bibitem[{Muraveva} et~al.(2025){Muraveva}, {Giannetti}, {Clementini}, {Garofalo}, and {Monti}]{2025MNRAS.536.2749M}
{Muraveva}, T.; {Giannetti}, A.; {Clementini}, G.; {Garofalo}, A.; {Monti}, L.
\newblock {Metallicity of RR Lyrae stars from the Gaia Data Release 3 catalogue computed with Machine Learning algorithms}.
\newblock {\em \mnras} {\bf 2025}, {\em 536},~2749--2769,  \href{http://arxiv.org/abs/2407.05815}{{\normalfont [arXiv:astro-ph.SR/2407.05815]}}.
\newblock {\url{https://doi.org/10.1093/mnras/stae2679}}.

\bibitem[Bhardwaj et~al.(2021)Bhardwaj, Rejkuba, Sloan, Marconi, and Yang]{Bhardwaj_2021}
Bhardwaj, A.; Rejkuba, M.; Sloan, G.C.; Marconi, M.; Yang, S.C.
\newblock Optical and Near-infrared Pulsation Properties of RR Lyrae and Population II Cepheid Variables in the Messier 15 Globular Cluster.
\newblock {\em The Astrophysical Journal} {\bf 2021}, {\em 922},~20.

\bibitem[Zgirski et~al.(2023)Zgirski, Pietrzy{\'n}ski, G{\'o}rski, Gieren, Wielg{\'o}rski, Karczmarek, Hajdu, Lewis, Chini, Graczyk, Ka{\l}uszy{\'n}ski, Narloch, Pilecki, Garc{\'\i}a, Suchomska, and Taormina]{Zgirski_2023}
Zgirski, B.; Pietrzy{\'n}ski, G.; G{\'o}rski, M.; Gieren, W.; Wielg{\'o}rski, P.; Karczmarek, P.; Hajdu, G.; Lewis, M.; Chini, R.; Graczyk, D.;  et~al.
\newblock New Near-infrared Period--Luminosity--Metallicity Relations for Galactic RR Lyrae Stars Based on Gaia EDR3 Parallaxes.
\newblock {\em The Astrophysical Journal} {\bf 2023}, {\em 951},~114.

\bibitem[Bhardwaj et~al.(2023)Bhardwaj, Marconi, Rejkuba, de~Grijs, Singh, Braga, Kanbur, Ngeow, Ripepi, Bono, De~Somma, and Dall'Ora]{Bhardwaj_2023}
Bhardwaj, A.; Marconi, M.; Rejkuba, M.; de~Grijs, R.; Singh, H.P.; Braga, V.F.; Kanbur, S.; Ngeow, C.C.; Ripepi, V.; Bono, G.;  et~al.
\newblock Precise Empirical Determination of Metallicity Dependence of Near-infrared Period--Luminosity Relations for RR Lyrae Variables.
\newblock {\em The Astrophysical Journal Letters} {\bf 2023}, {\em 944},~L51.

\bibitem[Narloch et~al.(2024)Narloch, Hajdu, Pietrzy{\'n}ski, Gieren, Zgirski, Wielg{\'o}rski, Karczmarek, G{\'o}rski, and Graczyk]{refId0}
Narloch, W.; Hajdu, G.; Pietrzy{\'n}ski, G.; Gieren, W.; Zgirski, B.; Wielg{\'o}rski, P.; Karczmarek, P.; G{\'o}rski, M.; Graczyk, D.
\newblock Period-luminosity and period-luminosity-metallicity relations for Galactic RR Lyrae stars in the Sloan bands.
\newblock {\em A\&A} {\bf 2024}, {\em 689},~A138.

\bibitem[{Gaia Collaboration}(2016)]{2016AA...595A...1G}
{Gaia Collaboration}.
\newblock {The Gaia mission}.
\newblock {\em \aap} {\bf 2016}, {\em 595},~A1,  \href{http://arxiv.org/abs/1609.04153}{{\normalfont [arXiv:astro-ph.IM/1609.04153]}}.
\newblock {\url{https://doi.org/10.1051/0004-6361/201629272}}.

\bibitem[{Gaia Collaboration} et~al.(2023){Gaia Collaboration}, {Vallenari}, {Brown}, {Prusti}, {de Bruijne}, {Arenou}, {Babusiaux}, {Biermann}, {Creevey}, {Ducourant}, {Evans}, {Eyer}, {Guerra}, {Hutton}, {Jordi}, {Klioner}, {Lammers}, {Lindegren}, {Luri}, {Mignard}, {Panem}, {Pourbaix}, {Randich}, {Sartoretti}, {Soubiran}, {Tanga}, {Walton}, {Bailer-Jones}, {Bastian}, {Drimmel}, {Jansen}, {Katz}, {Lattanzi}, {van Leeuwen}, {Bakker}, {Cacciari}, {Casta{\~n}eda}, {De Angeli}, {Fabricius}, {Fouesneau}, {Fr{\'e}mat}, {Galluccio}, {Guerrier}, {Heiter}, {Masana}, {Messineo}, {Mowlavi}, {Nicolas}, {Nienartowicz}, {Pailler}, {Panuzzo}, {Riclet}, {Roux}, {Seabroke}, {Sordo}, {Th{\'e}venin}, {Gracia-Abril}, {Portell}, {Teyssier}, {Altmann}, {Andrae}, {Audard}, {Bellas-Velidis}, {Benson}, {Berthier}, {Blomme}, {Burgess}, {Busonero}, {Busso}, {C{\'a}novas}, {Carry}, {Cellino}, {Cheek}, {Clementini}, {Damerdji}, {Davidson}, {de Teodoro}, {Nu{\~n}ez Campos}, {Delchambre}, {Dell'Oro}, {Esquej},
  {Fern{\'a}ndez-Hern{\'a}ndez}, {Fraile}, {Garabato}, {Garc{\'\i}a-Lario}, {Gosset}, {Haigron}, {Halbwachs}, {Hambly}, {Harrison}, {Hern{\'a}ndez}, {Hestroffer}, {Hodgkin}, {Holl}, {Jan{\ss}en}, {Jevardat de Fombelle}, {Jordan}, {Krone-Martins}, {Lanzafame}, {L{\"o}ffler}, {Marchal}, {Marrese}, {Moitinho}, {Muinonen}, {Osborne}, {Pancino}, {Pauwels}, {Recio-Blanco}, {Reyl{\'e}}, {Riello}, {Rimoldini}, {Roegiers}, {Rybizki}, {Sarro}, {Siopis}, {Smith}, {Sozzetti}, {Utrilla}, {van Leeuwen}, {Abbas}, {{\'A}brah{\'a}m}, {Abreu Aramburu}, {Aerts}, {Aguado}, {Ajaj}, {Aldea-Montero}, {Altavilla}, {{\'A}lvarez}, {Alves}, {Anders}, {Anderson}, {Anglada Varela}, {Antoja}, {Baines}, {Baker}, {Balaguer-N{\'u}{\~n}ez}, {Balbinot}, {Balog}, {Barache}, {Barbato}, {Barros}, {Barstow}, {Bartolom{\'e}}, {Bassilana}, {Bauchet}, {Becciani}, {Bellazzini}, {Berihuete}, {Bernet}, {Bertone}, {Bianchi}, {Binnenfeld}, {Blanco-Cuaresma}, {Blazere}, {Boch}, {Bombrun}, {Bossini}, {Bouquillon}, {Bragaglia}, {Bramante}, {Breedt},
  {Bressan}, {Brouillet}, {Brugaletta}, {Bucciarelli}, {Burlacu}, {Butkevich}, {Buzzi}, {Caffau}, {Cancelliere}, {Cantat-Gaudin}, {Carballo}, {Carlucci}, {Carnerero}, {Carrasco}, {Casamiquela}, {Castellani}, {Castro-Ginard}, {Chaoul}, {Charlot}, {Chemin}, {Chiaramida}, {Chiavassa}, {Chornay}, {Comoretto}, {Contursi}, {Cooper}, {Cornez}, {Cowell}, {Crifo}, {Cropper}, {Crosta}, {Crowley}, {Dafonte}, {Dapergolas}, {David}, {David}, {de Laverny}, {De Luise}, and {De March}]{2023AA...674A...1G}
{Gaia Collaboration}.; {Vallenari}, A.; {Brown}, A.G.A.; {Prusti}, T.; {de Bruijne}, J.H.J.; {Arenou}, F.; {Babusiaux}, C.; {Biermann}, M.; {Creevey}, O.L.; {Ducourant}, C.;  et~al.
\newblock {Gaia Data Release 3. Summary of the content and survey properties}.
\newblock {\em \aap} {\bf 2023}, {\em 674},~A1,  \href{http://arxiv.org/abs/2208.00211}{{\normalfont [arXiv:astro-ph.GA/2208.00211]}}.
\newblock {\url{https://doi.org/10.1051/0004-6361/202243940}}.

\bibitem[Riello et~al.(2021)Riello, De~Angeli, Evans, Montegriffo, Carrasco, Busso, Palaversa, Burgess, Diener, Davidson, Rowell, Fabricius, Jordi, Bellazzini, Pancino, Harrison, Cacciari, van Leeuwen, Hambly, Hodgkin, Osborne, Altavilla, Barstow, Brown, Castellani, Cowell, De~Luise, Gilmore, Giuffrida, Hidalgo, Holland, Marinoni, Pagani, Piersimoni, Pulone, Ragaini, Rainer, Richards, Sanna, Walton, Weiler, and Yoldas]{Riello_2021}
Riello, M.; De~Angeli, F.; Evans, D.W.; Montegriffo, P.; Carrasco, J.M.; Busso, G.; Palaversa, L.; Burgess, P.W.; Diener, C.; Davidson, M.;  et~al.
\newblock Gaia Early Data Release 3: Photometric content and validation.
\newblock {\em Astronomy \& Astrophysics} {\bf 2021}, {\em 649},~A3.
\newblock {\url{https://doi.org/10.1051/0004-6361/202039587}}.

\bibitem[{Jordi} et~al.(2010){Jordi}, {Gebran}, {Carrasco}, {de Bruijne}, {Voss}, {Fabricius}, {Knude}, {Vallenari}, {Kohley}, and {Mora}]{2010AA...523A..48J}
{Jordi}, C.; {Gebran}, M.; {Carrasco}, J.M.; {de Bruijne}, J.; {Voss}, H.; {Fabricius}, C.; {Knude}, J.; {Vallenari}, A.; {Kohley}, R.; {Mora}, A.
\newblock {Gaia broad band photometry}.
\newblock {\em \aap} {\bf 2010}, {\em 523},~A48.

\bibitem[{Gilligan} et~al.(2021){Gilligan}, {Chaboyer}, {Marengo}, {Mullen}, {Bono}, {Braga}, {Crestani}, {Dall'Ora}, {Fiorentino}, {Monelli}, {Neeley}, {Fabrizio}, {Mart{\'\i}nez-V{\'a}zquez}, {Th{\'e}venin}, and {Sneden}]{2021MNRAS.503.4719G}
{Gilligan}, C.K.; {Chaboyer}, B.; {Marengo}, M.; {Mullen}, J.P.; {Bono}, G.; {Braga}, V.F.; {Crestani}, J.; {Dall'Ora}, M.; {Fiorentino}, G.; {Monelli}, M.;  et~al.
\newblock {Metallicities from high-resolution spectra of 49 RR Lyrae variables}.
\newblock {\em \mnras} {\bf 2021}, {\em 503},~4719--4733,  \href{http://arxiv.org/abs/2103.11012}{{\normalfont [arXiv:astro-ph.SR/2103.11012]}}.
\newblock {\url{https://doi.org/10.1093/mnras/stab857}}.

\bibitem[{Green} et~al.(2019){Green}, {Schlafly}, {Zucker}, {Speagle}, and {Finkbeiner}]{2019ApJ...887...93G}
{Green}, G.M.; {Schlafly}, E.; {Zucker}, C.; {Speagle}, J.S.; {Finkbeiner}, D.
\newblock {A 3D Dust Map Based on Gaia, Pan-STARRS 1, and 2MASS}.
\newblock {\em \apj} {\bf 2019}, {\em 887},~93,  \href{http://arxiv.org/abs/1905.02734}{{\normalfont [arXiv:astro-ph.GA/1905.02734]}}.
\newblock {\url{https://doi.org/10.3847/1538-4357/ab5362}}.

\bibitem[{Schlegel} et~al.(1998){Schlegel}, {Finkbeiner}, and {Davis}]{1998ApJ...500..525S}
{Schlegel}, D.J.; {Finkbeiner}, D.P.; {Davis}, M.
\newblock {Maps of Dust Infrared Emission for Use in Estimation of Reddening and Cosmic Microwave Background Radiation Foregrounds}.
\newblock {\em \apj} {\bf 1998}, {\em 500},~525--553,  \href{http://arxiv.org/abs/astro-ph/9710327}{{\normalfont [arXiv:astro-ph/astro-ph/9710327]}}.
\newblock {\url{https://doi.org/10.1086/305772}}.

\bibitem[{Cardelli} et~al.(1989){Cardelli}, {Clayton}, and {Mathis}]{1989ApJ...345..245C}
{Cardelli}, J.A.; {Clayton}, G.C.; {Mathis}, J.S.
\newblock {The Relationship between Infrared, Optical, and Ultraviolet Extinction}.
\newblock {\em \apj} {\bf 1989}, {\em 345},~245.
\newblock {\url{https://doi.org/10.1086/167900}}.

\bibitem[{Lomb}(1976)]{1976ApSS..39..447L}
{Lomb}, N.R.
\newblock {Least-Squares Frequency Analysis of Unequally Spaced Data}.
\newblock {\em \apss} {\bf 1976}, {\em 39},~447--462.
\newblock {\url{https://doi.org/10.1007/BF00648343}}.

\bibitem[{Scargle}(1982)]{1982ApJ...263..835S}
{Scargle}, J.D.
\newblock {Studies in astronomical time series analysis. II. Statistical aspects of spectral analysis of unevenly spaced data.}
\newblock {\em \apj} {\bf 1982}, {\em 263},~835--853.
\newblock {\url{https://doi.org/10.1086/160554}}.

\bibitem[{Simon} and {Lee}(1981)]{1981ApJ...248..291S}
{Simon}, N.R.; {Lee}, A.S.
\newblock {The structural properties of cepheid light curves.}
\newblock {\em \apj} {\bf 1981}, {\em 248},~291--297.
\newblock {\url{https://doi.org/10.1086/159153}}.

\bibitem[{Deb} and {Singh}(2009)]{deb2009}
{Deb}, S.; {Singh}, H.P.
\newblock {Light curve analysis of variable stars using Fourier decomposition and principal component analysis}.
\newblock {\em \aap} {\bf 2009}, {\em 507},~1729--1737,  \href{http://arxiv.org/abs/0903.3500}{{\normalfont [arXiv:astro-ph.SR/0903.3500]}}.
\newblock {\url{https://doi.org/10.1051/0004-6361/200912851}}.

\bibitem[{Deb} and {Singh}(2010)]{2010MNRAS.402..691D}
{Deb}, S.; {Singh}, H.P.
\newblock {Physical parameters of the Small Magellanic Cloud RR Lyrae stars and the distance scale}.
\newblock {\em \mnras} {\bf 2010}, {\em 402},~691--704.

\bibitem[{Li} et~al.(2023){Li}, {Huang}, {Liu}, {Beers}, and {Zhang}]{2023ApJ...944...88L}
{Li}, X.Y.; {Huang}, Y.; {Liu}, G.C.; {Beers}, T.C.; {Zhang}, H.W.
\newblock {Photometric Metallicity and Distance Estimates for 136,000 RR Lyrae Stars from Gaia Data Release 3}.
\newblock {\em \apj} {\bf 2023}, {\em 944},~88,  \href{http://arxiv.org/abs/2206.07668}{{\normalfont [arXiv:astro-ph.SR/2206.07668]}}.
\newblock {\url{https://doi.org/10.3847/1538-4357/acadd5}}.

\bibitem[{Layden}(1994)]{1994AJ....108.1016L}
{Layden}, A.C.
\newblock {The Metallicities and Kinematics of RR Lyrae Variables. I. New Observations of Local Stars}.
\newblock {\em \aj} {\bf 1994}, {\em 108},~1016.
\newblock {\url{https://doi.org/10.1086/117132}}.

\bibitem[{Clementini} et~al.(2019){Clementini}, {Ripepi}, {Molinaro}, {Garofalo}, {Muraveva}, {Rimoldini}, {Guy}, {Jevardat de Fombelle}, {Nienartowicz}, {Marchal}, {Audard}, {Holl}, {Leccia}, {Marconi}, {Musella}, {Mowlavi}, {Lecoeur-Taibi}, {Eyer}, {De Ridder}, {Regibo}, {Sarro}, {Szabados}, {Evans}, and {Riello}]{2019AA...622A..60C}
{Clementini}, G.; {Ripepi}, V.; {Molinaro}, R.; {Garofalo}, A.; {Muraveva}, T.; {Rimoldini}, L.; {Guy}, L.P.; {Jevardat de Fombelle}, G.; {Nienartowicz}, K.; {Marchal}, O.;  et~al.
\newblock {Gaia Data Release 2. Specific characterisation and validation of all-sky Cepheids and RR Lyrae stars}.
\newblock {\em \aap} {\bf 2019}, {\em 622},~A60,  \href{http://arxiv.org/abs/1805.02079}{{\normalfont [arXiv:astro-ph.SR/1805.02079]}}.
\newblock {\url{https://doi.org/10.1051/0004-6361/201833374}}.

\bibitem[{Zinn} and {West}(1984)]{1984ApJS...55...45Z}
{Zinn}, R.; {West}, M.J.
\newblock {The globular cluster system of the Galaxy. III. Measurements of radial velocity and metallicity for 60 clusters and a compilation of metallicities for 121 clusters.}
\newblock {\em \apjs} {\bf 1984}, {\em 55},~45--66.
\newblock {\url{https://doi.org/10.1086/190947}}.

\bibitem[{Carretta} et~al.(2009){Carretta}, {Bragaglia}, {Gratton}, {D'Orazi}, and {Lucatello}]{2009AA...508..695C}
{Carretta}, E.; {Bragaglia}, A.; {Gratton}, R.; {D'Orazi}, V.; {Lucatello}, S.
\newblock {Intrinsic iron spread and a new metallicity scale for globular clusters}.
\newblock {\em \aap} {\bf 2009}, {\em 508},~695--706,  \href{http://arxiv.org/abs/0910.0675}{{\normalfont [arXiv:astro-ph.GA/0910.0675]}}.
\newblock {\url{https://doi.org/10.1051/0004-6361/200913003}}.

\bibitem[{D{\'e}k{\'a}ny} et~al.(2021){D{\'e}k{\'a}ny}, {Grebel}, and {Pojma{\'n}ski}]{2021ApJ...920...33D}
{D{\'e}k{\'a}ny}, I.; {Grebel}, E.K.; {Pojma{\'n}ski}, G.
\newblock {Metallicity Estimation of RR Lyrae Stars From Their I-Band Light Curves}.
\newblock {\em \apj} {\bf 2021}, {\em 920},~33,  \href{http://arxiv.org/abs/2107.05983}{{\normalfont [arXiv:astro-ph.SR/2107.05983]}}.
\newblock {\url{https://doi.org/10.3847/1538-4357/ac106f}}.

\bibitem[{Liu} et~al.(2022){Liu}, {Huang}, {Bird}, {Zhang}, {Wang}, and {Tian}]{2022MNRAS.517.2787L}
{Liu}, G.; {Huang}, Y.; {Bird}, S.A.; {Zhang}, H.; {Wang}, F.; {Tian}, H.
\newblock {Probing the Galactic halo with RR lyrae stars - III. The chemical and kinematic properties of the stellar halo}.
\newblock {\em \mnras} {\bf 2022}, {\em 517},~2787--2800,  \href{http://arxiv.org/abs/2209.07885}{{\normalfont [arXiv:astro-ph.GA/2209.07885]}}.
\newblock {\url{https://doi.org/10.1093/mnras/stac2666}}.

\bibitem[{Jurcsik} and {Hajdu}(2023)]{2023MNRAS.525.3486J}
{Jurcsik}, J.; {Hajdu}, G.
\newblock {Photometric metallicities of fundamental-mode RR Lyr stars from Gaia G band photometry of globular-cluster variables}.
\newblock {\em \mnras} {\bf 2023}, {\em 525},~3486--3498,  \href{http://arxiv.org/abs/2308.08929}{{\normalfont [arXiv:astro-ph.SR/2308.08929]}}.
\newblock {\url{https://doi.org/10.1093/mnras/stad2510}}.

\end{thebibliography}
\end{document}